\documentclass[%
 reprint,
superscriptaddress,
 amsmath,amssymb,
 aps,
 prl
]{revtex4-1}

\usepackage{graphicx}
\usepackage{dcolumn}
\usepackage{bm}
\usepackage{color}

\newcommand\Ca{\text{Ca}}

\newcommand\mean[1]{\left\langle #1 \right\rangle}

\usepackage{natbib}

\begin{document}

\preprint{APS/123-QED}

\title{Yield stress fluid behavior of foam in porous media}

\author{Alexis Mauray}
\affiliation{Univ. Grenoble Alpes, CNRS, Grenoble-INP, Lab. LRP UMR 5520, F-38000 Grenoble, France}
\author{Max Chabert}
\affiliation{Solvay, 1 Biopolis Drive, 138622 Singapore.}
\author{Hugues Bodiguel}%
 \email{hugues.bodiguel@univ-grenoble-alpes.fr}
\affiliation{Univ. Grenoble Alpes, CNRS, Grenoble-INP, Lab. LRP UMR 5520, F-38000 Grenoble, France}

\begin{abstract}
We report a local analysis of the flow of foams in two dimensional heterogeneous model porous media. Pressure measurements are combined with direct observation in order to determine simultaneously the effective viscosity and the fraction of trapped gas. Experiments 
are conducted in the low-quality regime, in a large range of capillary numbers and relative gas flow rates. The effective viscosity is very high (up to a few thousands times that of water) at low capillary numbers and rheology can be described 
by a decreasing power-law function of the capillary number of exponent ranging from -1 to -0.75. The flow is heterogeneous and the fraction of preferential paths increases with relative gas flow rate. However, increase in fraction of preferential paths remains very limited for increasing capillary number, thus failing to account for the strongly shear thinning behavior observed. Additional experiments motivated by this finding, carried out in a straight channel of varying cross-section, show a deviation from Bretherton law at low capillary numbers in this geometry with the bubble train behaving as a shear-thinning yield stress fluid. The pressure threshold evidenced accounts quantitatively for the effective viscosity data in the micromodel, indicating that it is the main mechanism underlying the measured foam rheology.

\end{abstract}

\pacs{Valid PACS appear here}
 
\maketitle


Many industrial processes involve gas/liquid flows through porous media. Prominent examples can be found in petroleum~\cite{lake} and in chemical engineering~\cite{al-dahhan_high-pressure_1997}. Some take advantage of the addition of surfactants to stabilize the gas-liquid interfaces leading to formation of foam into earth' subsurface flows, in particular to enhance oil recovery or to clean polluted soils~\cite{Mulligan}. Injecting foams into rocks present numerous advantages over water flooding as the needed volumes of water are considerably lowered and as the specific rheological properties of the foam allow a better invasion into small pores~\cite{a._r._kovscek_fundamentals_1993}. Foam also improves macroscopic sweep efficiency because of its high effective viscosity which reduces both channeling flows (flows of gas in the high permeability streaks) and viscous fingering. In this context, the understanding of physical mechanisms behind foam flow behavior in porous media has been subject of many studies and still holds open questions as of today.

\paragraph*{}
Most experimental studies focused on macroscopic flows of foam in realistic porous materials such as packed grains or porous solids (\cite{z._i._khatib_effects_1988,Bertin,j._m._alvarez_unified_2001,tang_trapped_2006}). Although these studies provide important insights, flow at the level of the pore cannot be accessed, and they only yield a limited microscopic description of the mechanisms ruling foam flows in porous media. One of the major difficulty originates from the multiple physical mechanisms at work \cite{a._r._kovscek_fundamentals_1993} (foam formation and lamella division, coalescence, transport, channeling,...) and the multiple parameters involved, usually not accessible experimentally in real rocks.

\paragraph*{}
Therefore, two-dimensional micromodels have been used for a few decades to better understand the relevant mechanisms in two phase flows in porous media \cite{Lenormand_prl1985,Frette_prl1985,cottin_drainage_2010}. Such devices have been used in the past few years to visualize foam flows in heterogeneous media \cite{ma_visualization_2012,conn_visualizing_2014,gauteplass_pore-level_2015}, but have not yet been applied to quantitatively characterize the effective viscosity of foam in porous media, although it appears as foam central attribute from the application standpoint.. Bubble and droplet traffic has been extensively studied and is rather well understood in the dilute limit pertaining to applications in digital microfluidics \cite{engl_droplet_2005,Choi_2011,fuerstman_pressure_2007,parthiban_bistability_2013,Champagne_2010,hourtane_dense_2016}, but these studies focus on too simple geometries or on too dilute regimes for being directly extended to porous media.

The high effective viscosity of foam in porous medium - interchangeably referred to as mobility reduction, exhibits two regimes macroscopically, depending on the relative gas flow rate $f_g$ \cite{osterloh_effects_1992,alvarez_unified_2001}. In the low quality regime, and below a critical value $f_g \sim 0.8 $, it increases when $f_g$ is increased \cite{lotfollahi_comparison_2016}, whereas at higher $f_g$, it decreases due to lamella coalescence. In the low quality regime, on which we focus here, recent experimental results reveal that the foam is strongly shear-thinning; its viscosity scales as $\Ca^{-n}$, with $n$ ranging from 0.6 to 0.9 \cite{gassara_calibrating_2017,jones_foam_2018}, $\Ca$ being the capillary number, defined as $\Ca=\mu V/\gamma$, where $\mu$ is the water viscosity, $V$ the velocity and $\gamma$ the surface tension. The high effective viscosity is usually described in mechanistic models as the combination of two mechanisms~\cite{kovscek_fundamentals_nodate,hematpur_foam_2018}. First, the motion of the dispersed gas phase involves dynamic menisci~\cite{hirasaki_mechanisms_1985,hourtane_dense_2016}, as theoretically studied by Bretherton \cite{bretherton_motion_1961}, which induces extra dissipation that scales as $\Ca^{2/3}$. Second, as inferred \cite{kovscek_fundamentals_nodate} and experimentally demonstrated \cite{tang_trapped_2006,gauteplass_pore-level_2015}, only a fraction of the gas contributes to the flow in preferential paths, while the other fraction remains trapped. Intuitatively, it is tempting to attribute the strong shear thinning behavior to the combination of the two mechanisms. We experimentally demonstrate here that none of them are relevant. 

We consider in this work foam flow in a heterogeneous 2D porous medium of well-controlled geometry. The key experimental results are obtained by combining pressure measurements and direct observations, varying both gas and liquid flow rates in a very large range. We observe the generic behavior and mobility reduction values that have been reported in 3D porous media, in the low quality regime. In particular, we find that the mobility reduction scales as a power law of the capillary number: Ca$^{-n}$, with $n$ between 0.7 and 0.9. In contrast, unexpectedly, the number of preferential paths exhibits a weak dependency with $\Ca$, ruling out their impact on foam shear thinning behavior. We then report results from further experiments prompted by this finding. They consists in measurements in a simpler geometry, i.e. in a single channel of varying cross-section, and evidence a deviation from Bretherton law at low $\Ca$. Using this result, we are able to account quantitatively for the mobility reduction measured in the micromodel.

\begin{figure}[h!]
\centering
\includegraphics[width=0.92\linewidth]{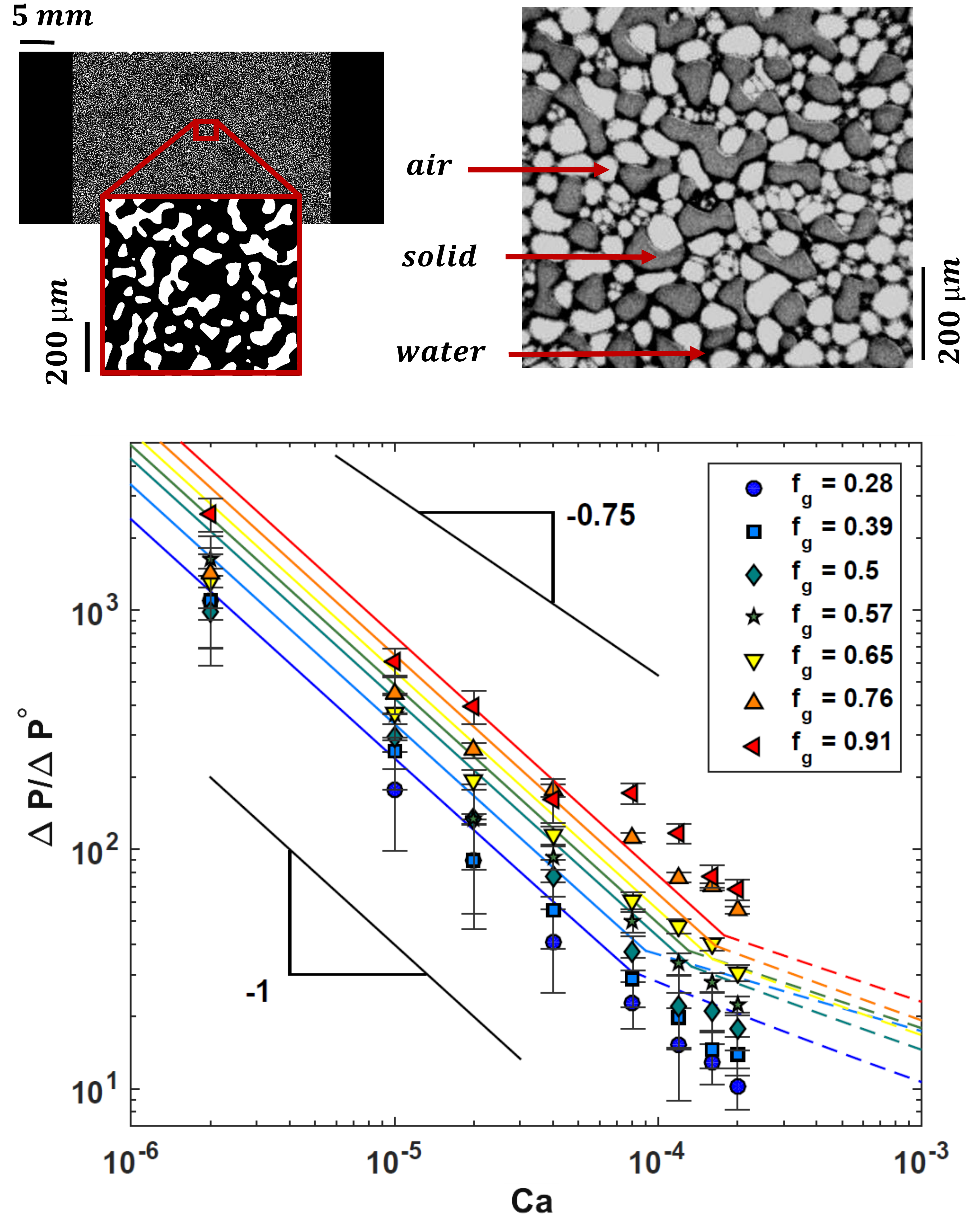}
\caption{\textit{Top:} images of the mask used to fabricate the device and magnified image taken during foam flow. \textit{Bottom: }Mobility reduction $\Delta P / \Delta P^{\circ}$ as function of $\Ca$ for various $f_g$. Lines correspond to the semi-empiric predictions (see text): solid ones for the threshold regime, dashed ones for the Bretherton regime. }
\centering
\label{fig:mob}
\end{figure}

The device used in this study is a NOA microfluidic chip consisting in a $L\times W = 3 \times 2$ cm 2D random porous medium, obtained using a simple numerical algorithm (see supplemental materials for details~\cite{supmat}, section IB), and displayed in Fig.~\ref{fig:mob}, top-left. The height $h$ of the device is 180$\mu$m. The correlation length of the porous medium is set to $l_{c}=200$ $\mu$m, its 2D porosity $\phi$ to 0.7. The pore size exhibits a mean value of $\mean{w}=134 \mu$m while the mean pore throat diameter is about 95 $\mu$m. Its permeability has been determined experimentally, $k=9.25\times10^{-11}m^2$ (see supplemental materials for details~\cite{supmat}, section IC). 

The flow is controlled using a syringe pump (Nemesys), and the outlet is connected to a pressurized reservoir, set at 3 bars to allow neglecting impact of gas compressibility. A differential pressure sensor is used between the inlet and the outlet to determine the pressure drop. The experimental setup is complemented by a fast camera (Mikrotron) mounted on a optical lens (Nikon). All experiments are carried out with an aqueous solution of 5$\%$ in weight of Solvay \textit{SurfEOR} surfactant that allows avoiding bubble coalescence and ensures total wetting. The surface tension $\gamma$ of the solution is $33$ mN.m$^{-1}$, as determined using the pendant drop method. 

The gas-liquid co-injection protocol has been carefully established to avoid entrance effects and to dissociate foam formation from its steady state flow properties. We proceed as follows: first, we generate big bubbles by using a T-junction connected to two syringes, controlled at flow rates of $Q_g$ and $Q_l$, for the gas and liquid phase, respectively. When entering the porous medium, these bubbles progressively divide provided that the total flow rate is high enough. Eventually they reach a size which is similar to that of the pore size, so that division mechanisms are subsequently inefficient \cite{chen_insights_2005,gauteplass_pore-level_2015}. Then, in order to ensure a steady state, we collect the effluents and reinject them into the porous domain once again at high flow rate. In this second step, we observe that the mean bubble size remains constant. We thus obtain a "`foam"' which consists in bubbles of about the pore size, generated by the porous medium itself. This foam is finally injected at a desired flow rate, which is generally much lower. By construction, the relative gas flow rate $f_g=Q_g/(Q_g+Q_l)$ is controlled by the first generation step. This protocol ensures that the injected phases have the same texture whatever the value of $\Ca$. In the following, the capillary number is precisely defined as $\Ca=\mu \left(Q_l+Q_g\right)/ \gamma \phi h W$, where $\mu$ is the aqueous solution viscosity ($(1 \pm 0.08) \times 10^{-3}$ Pa.s).

Figure \ref{fig:mob}(bottom) summarizes the main result of this work. It displays the mobility reduction, i.e. the measured pressure drop $\Delta P$ in presence of bubbles normalized by the pressure drop $\Delta P^{\circ}$ due to water flowing at the same total flow rate. For every $f_g$ tested, the mobility reduction decreases when the capillary number is increased, and follows a power law of exponent ranging between -1 for the lowest $f_g$ tested and -0.75 for the highest one. When $f_g$ is increased, the effective viscosity increases as well. Let us highlight that the very high values of the mobility reductions reached at low $\Ca$ (more than 2000!) correspond to orders of magnitude reported for foams flows in real rocks \cite{z._i._khatib_effects_1988}. Furthermore, the power law behavior with respect to $\Ca$ is similar to that in real porous media \cite{lotfollahi_comparison_2016,gassara_calibrating_2017,jones_foam_2018}. Thus, these model experiments contain the minimal physical ingredients for describing mobility reduction in foam flow in porous media. 

The movies acquired together with the pressure measurements reveal that the flow is heterogeneous (see for example the movies available in the supplemental materials \cite{supmat}): in some regions of the porous domain, gas bubbles are trapped and do not contribute to the flow. The latter is concentrated in preferential paths. In order to characterize these, we compute the absolute difference between successive images and average in time (see supplemental materials for details \cite{supmat}). Examples are displayed in Fig.~\ref{fig:ch}, and the complete data set is available in the supplemental materials~\cite{supmat}. These images highlight qualitatively the preferential paths. One can notice that their number slightly increases when increasing $\Ca$, but also when increasing $f_g$. More quantitatively, we estimate the fraction of the pores that contributes to the flow by computing the mean value of the binary image obtained after thresholding. This fraction is defined thereafter as the ratio of effective flowing paths $N$ over the number of available ones $N_0$. The results are displayed in Fig.~\ref{fig:ch}. For a given $f_g$, $N/N_0$ does not greatly vary when varying $\Ca$. On the contrary, this ratio strongly depends on $f_g$. Strikingly, the higher $f_g$ is, the more homogeneous the flow is.

Since we know precisely the characteristics of the porous medium and we get an estimate of the ratio of preferential paths, we now try to discuss more quantitatively the mobility reduction values. The simplest model to test consists in a bundle of $N$ parallel channels, in which bubble trains are flowing. Dedicated microfluidic studies \cite{hourtane_dense_2016} on bubble trains have verified that Bretherton law~\cite{bretherton_motion_1961} applies, even for dense bubble trains. Each bubble creates an extra pressure drop proportional to $p_c \Ca_l^{2/3}$, where $p_c$ is the mean capillary pressure, set by the channel geometry: $p_c=2\gamma\left(1/h + 1/ \langle w \rangle \right)$, and $\Ca_l$ the local capillary number. Since only a fraction $N/N_0$ of the medium contributes to the flow, local velocities are greater by a factor of $N_0/N$ than the pore velocity estimated for a mean homogeneous flow, i.e. $\Ca_l=\Ca N_0/N$. Neglecting the other contributions to the total pressure drop - this is justified by the high values of the mobility reduction -, we expect that the latter would be proportional to $\Ca^{-1/3} \left(N_0/N\right)^{2/3}$, since $\Delta P_0$ is proportional to $\Ca$. However, as the fraction of preferential paths only shows a weak dependency with $\Ca$, the above model cannot account for the strong shear-thinning found experimentally. We conclude that Bretherton law in straight channels cannot be directly extrapolated to the tortuous flow paths in the porous medium.

\begin{figure}[h!]
\centering
\includegraphics[width=0.9\linewidth]{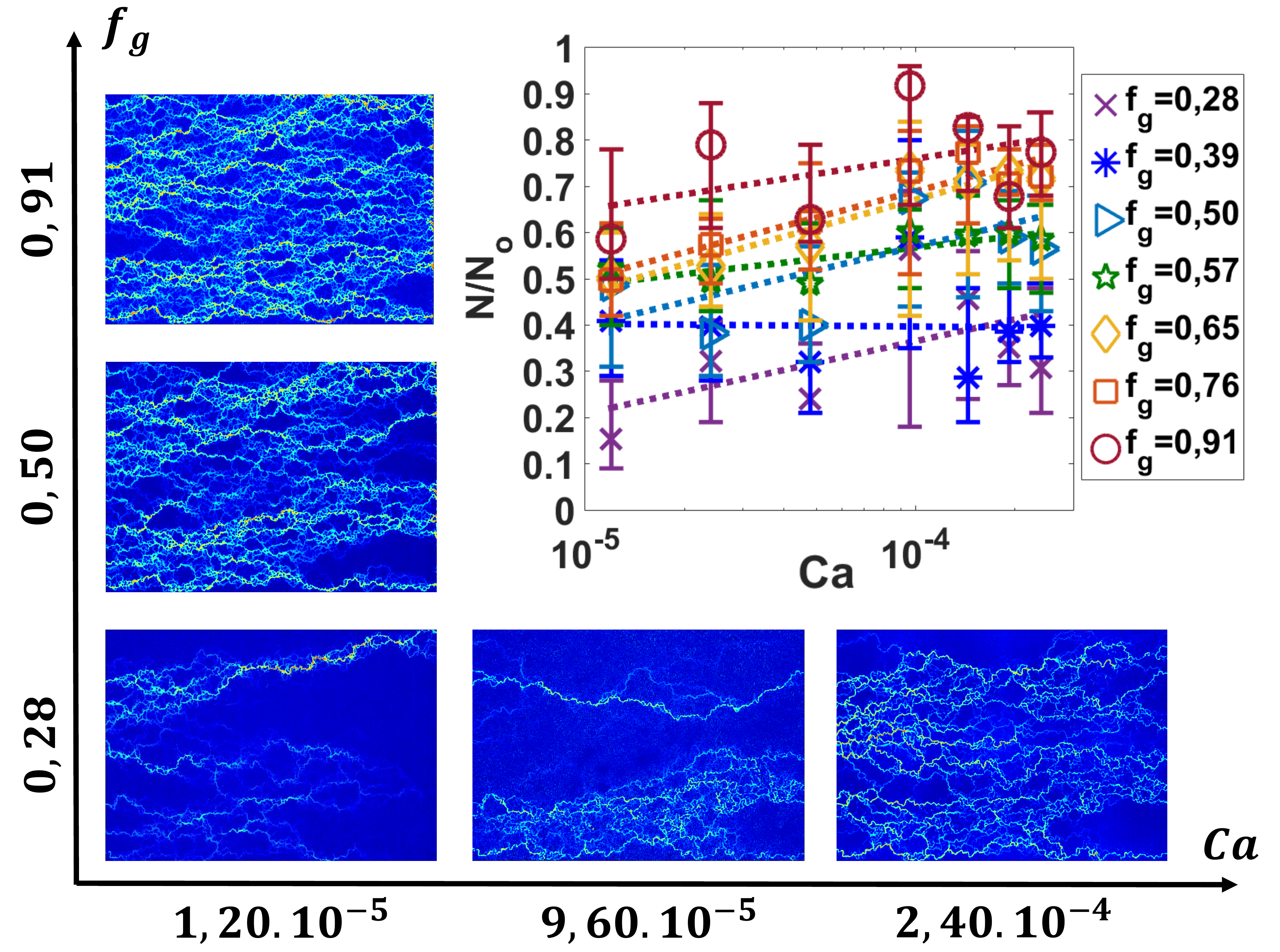}
\caption{Examples of time averaged image differences highlighting preferential paths. In insert, the fraction of these paths, $N/N_0$, is plotted as a function of $\Ca$ for various relative gas flow rates $f_g$.}
\centering
\label{fig:ch}
\end{figure}

In order to get deeper insights on the mechanisms responsible for the mobility reduction, we performed similar experiments in two much simpler geometries, which consist in a single channel, either of uniform cross-section, either of sinusoidal cross-section. In the second case, our intention is to mimic the succession of constrictions during bubble motion in the porous medium. The geometric features of this device (width and height, amplitude and wavelength of the sine function) are close to that of the pores in the micromodel. Details are given in the supplemental materials~ \cite{supmat}. Using a flow focusing element~\cite{anna_formation_2003}, we inject a periodic train of monodisperse bubble in these channels. The bubble size is slightly larger than the mean width. We systematically measure the pressure drop as a function of the two parameters of the problem: the wavelength of the bubble train and the capillary number. 

\begin{figure}[h!]
\centering
\includegraphics[scale=0.28]{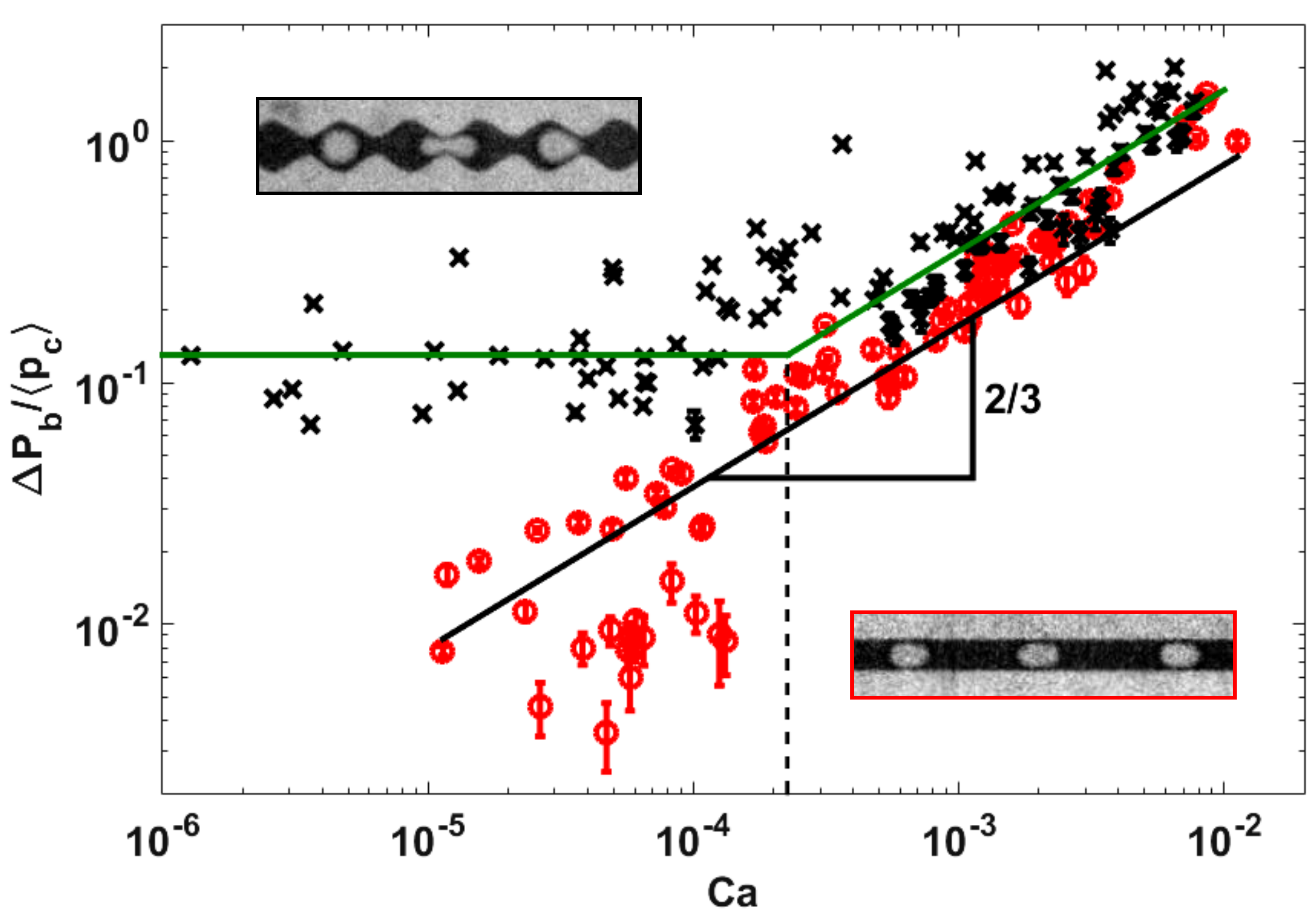}
\caption{Pressure drop induced by a single bubble $\Delta P_b$, normalized by the capillary pressure $\mean{p_c}=2\gamma\left(1/h+1/\mean{w}\right)$ as a function of $\Ca$, in two geometries: a channel of uniform cross-section (red circles) and a channel of sinusoidal width (black crosses). Data correspond to various wavelengths, so that this parameter appears irrelevant. Solid lines represent the best fit to the data. For the uniform cross-section case, we find a Bretherton law $\beta\Ca^{2/3}$ with $\beta=17$. For the varying cross-section case, we find $\beta=35$ in the Bretherton regime (for $\Ca>2.3\times 10^{-4}$) and $\alpha=0.13$ in the threshold regime. } 
\centering
\label{fig:bretherton}
\end{figure}

We aim at measuring the extra pressure drop due to a single bubble, $\Delta P_b$. For that purpose, we substract from the total pressure the contribution of the continuous phase in between the bubbles, assuming Poiseuille law, and then divide it by the number of bubbles in the channels estimated knowing the wavelength of the bubble train \cite{fuerstman_pressure_2007,hourtane_dense_2016}. The results are displayed in Fig.~\ref{fig:bretherton}. For both channels, all the data collapse when plotted as a function of $\Ca$. This shows that the bubble density has no direct influence on pressure drop \textit{per} bubble, though the total pressure drop depends on the bubble wavelength. For the channel of uniform cross-section, the data are rather well accounted by Bretherton law, $\Delta P_b = \alpha p_c \Ca^{2/3}$. The empiric prefactor $\alpha$ is quite high - and different from the theoretical prediction of Bretherton- since we find $\alpha = 17$. This deviation is similar to previously reported data \cite{fuerstman_pressure_2007, hourtane_dense_2016} and might be associated to the impact of the surfactant. For the channel of sinusoidal cross-section, we find a very different behavior. Two regimes are evidenced around a threshold capillary number $\Ca_c = 2.3 \times 10^{-4}$. For $\Ca>\Ca_c$, we recover a Bretherton-like law, with $\alpha\simeq 35$. At lower capillary numbers, data deviate from this law and the bubble pressure drop exhibits a plateau, around $\beta\mean{ p_c }$, where $\beta \simeq 0.13$. The motion of a bubble train in a channel with constrictions is thus similar to that of a yield stress fluid: there is a pressure threshold below which a bubble does not flow, and its value is a fraction of the capillary pressure. In the following, we refer to these two regimes as the \textit{threshold regime} and the \textit{Bretherton regime}.

Since the geometrical features of the channel of sinusoidal cross-section are very similar to the ones of the porous media, we may use quantitatively the results obtained in the channel and extrapolate to the porous medium case. By estimating the number of bubbles involved in one of the preferential path, we obtain (see supplemental materials for details~\cite{supmat}) that, in the threshold regime, $\Delta P/\Delta P_0 = a \beta f_g \Ca^{-1}$, where the geometric constant $a \simeq 0.066$. When the local capillary number exceeds $\Ca_c$, i.e. for $\Ca N_0/N > \Ca_c$, the Bretherton regime should hold and we expect that $\Delta P/\Delta P_0 = a \alpha f_g \left(N_0/N\right)^{2/3} \Ca^{-1/3}$. 

The two regimes are plotted in Figure \ref{fig:mob}. It turns out that most experiments fall in the threshold regime, only the highest tested flow rate being in the Bretherton regime. Thus, the present set of data does not evidence the transition. Nevertheless, the above semi-empiric predictions accounts rather nicely for the mobility reduction data, although they slightly overestimate the mobility reduction at high $f_g$ and low $\Ca$. We can thus infer that the threshold regime also occurs in the porous medium, and that it is responsible for the strong apparent shear-thinning effect which is measured. Strikingly, the number of preferential paths has no direct effect on the mobility reduction in this regime and only influences the onset of the Bretherton regime. 

\begin{figure}[h!]
\centering
\includegraphics[width=\linewidth]{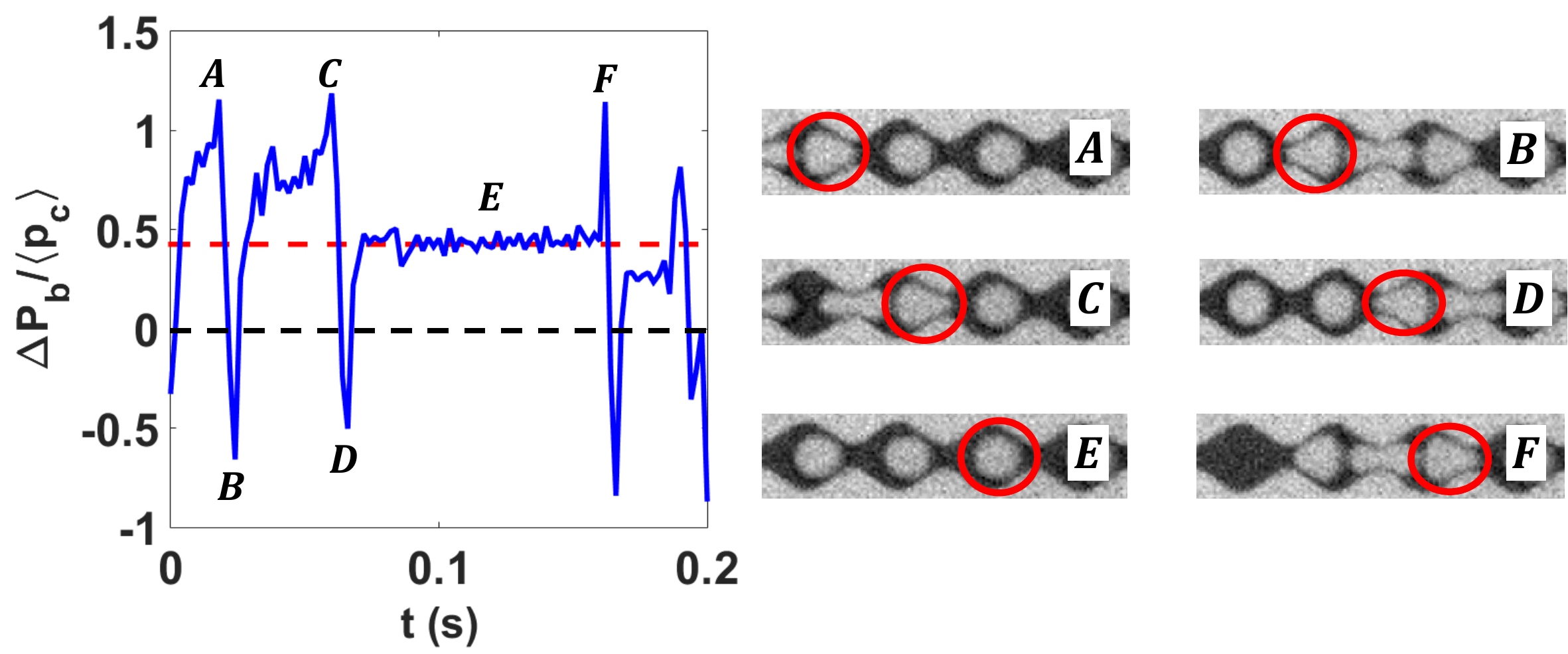}
\caption{Example of normalized instantaneous pressure difference per bubble as a function of time, as determined by the estimation of the menisci curvature on the images. In this experiment, $\Ca = 10^{-5}$. In the 6 images labeled from A to F, the bubble of interest is highlight in red. The time corresponding to this six images are displayed on the plot. The dashed line represents the mean value. }
\centering
\label{fig:bubblemotion}
\end{figure}

Eventually, it is worth discussing the physical mechanisms underlying the threshold regime. Such a threshold has been predicted theoretically \cite{sinha_effective_2013}. It originates from a coupling between the meniscus motion and the capillary pressure difference that appears between the front and rear menisci of a bubble passing a constriction. Although this pressure difference is symmetric with respect to the bubble position, the fact that a bubble moves slower upstream than downstream the constriction leads after averaging over time and space to a net pressure drop. It is of the order of a fraction of $\left\langle p_c \right\rangle $, as observed in our experiments. To verify this mechanism, we perform image analysis on the movie acquired in the straight channel of varying cross-section, and estimate the curvature difference between the front and rear menisci of single bubbles. As shown in Fig. \ref{fig:bubblemotion}, we find that it fluctuates in correspondence with bubble displacement through successive constrictions. Bubble motion is non-steady and the bubble spends more time upstream of the constriction where the pressure difference between front and rear menisci is positive rather downstream where it is negative. As a result, pressure mean value is non-zero and in qualitative agreement with the value of $\beta=0.13$ obtained from the pressure drop measurement. 

In summary, we report in this letter a complete data set of foam apparent viscosities in the low-quality regime, acquired in a bidimensional transparent porous medium. We evidence, as in real rocks, high values of apparent viscosity and a strongly shear thinning behavior. Direct observation reveals that the flow is heterogeneous, and concentrated into some preferential paths. Interestingly, the flow becomes more homogeneous when the bubble density is high. This result has strong practical consequences, since foam is generally used (either in oil recovery or in soil remediation) to better sweep porous media and the most homogeneous flow is interesting for that purpose. Contradicting an usual paradigm, we show that variation in number of preferential paths with capillary number is irrelevant to adequately describe foam rheology in porous media. Indeed, using model experiments in a simple channel of varying cross-section, we evidence that, at the low capillary numbers investigated, the pressure drop per bubble actually becomes independent on the flow rate and is simply a fraction of the capillary pressure. Extrapolation of this result to the porous medium allows, without additional significant assumptions, to capture reasonably well the measured values of the mobility reduction in a large range of $\Ca$ and $f_g$. Some features of the problem still remain to be investigated. We observed, but did not analyze in details, that the motion of the bubbles is locally not steady but intermittent at high bubble densities, even though the flow appear rather homogeneous when averaged over time. This behavior may explain that at high relative gas flow rate, the shear thinning exponent of the mobility reduction is about 0.75, whereas it is closer to 1 at lower $f_g$. This progressive variation of the exponent is currently not captured by the model, and ask for refinements and additional work. It would also be interesting to increase further the capillary number in order to confirm the existence of a Bretherton-like regime above $\Ca_c$.

\begin{acknowledgments}
The authors acknowledge Solvay for financial support. LRP is a part of the LabEx Tec 21 (Investissements d'Avenir, Grant Agreement No. ANR-11-LABX-0030) and PolyNat Carnot Institute (ANR-11-CARN-030-01). 

\end{acknowledgments}


\end{document}